\documentclass{ptephy}    
\usepackage{bm}
\usepackage[]{graphicx}
\usepackage{here}
\usepackage{caption}
\usepackage{tabularx}
\usepackage{amsmath,amssymb}
\usepackage{wrapfig}
\usepackage{color}
\usepackage{braket}
\usepackage{enumerate}

\begin{document}
\title{Corner functions from entanglement indices of harmonic lattices}
\author{\name{\fname{Masafumi} \surname{Shimojo}}{1,\ast}, \name{\fname{Satoshi} \surname{Ishihara}}{2}, 
\name{\fname{Hironobu} \surname{Kataoka}}{2}, \name{\fname{Atsuko} \surname{Matsukawa}}{2}\\ 
}

\address{
\affil{1}{Department of Electronics and Information Engineering, National Institute of Technology, Fukui College,
Geshicho, Sabae, Fukui 916-8507, Japan}
\affil{2}{Department of Physics, Hyogo University of Education, Shimokume, Kato, Hyogo 673-1494, Japan}
\email{satoshi@yukawa.kyoto-u.ac.jp}
}
\begin{abstract}%
We study the entanglement indices such as logarithmic negativities (LNs) and mutual informations (MIs) between two adjacent subsets in an isolated universal set $U$ of harmonic oscillators arranged on a two dimensional lattice within a sufficiently large square. First, we verify the values of the corner functions of angle $\pi/2, \pi/4, 3\pi/4$ presented in the previous study which adopts periodic boundary conditions (PBCs) for the $U$. The values of each corner function obtained from LNs are nearly equal to those in the previous ones, while those of $3\pi/4$ from  MIs are not sufficiently consistent with those computed from LNs. Next, for the case where the universal system $U$ satisfies the fixed boundary conditions(FBCs), we calculate LNs, MIs at several locations in $U$, compare them, especially corner functions with the values obtained in the PBCs case, and examine the effect of the fixed ends. In addition, we examine Renyi entropies for sets of three dimensional lattice sites, the corner functions and the edge terms with solid angles and dihedral angles $\pi/2,\pi/4$, respectively.   
\end{abstract}
\subjectindex{A60,A61}
\maketitle
\section{Introduction}
  Entanglement entropy which represents the measure of the loss of quantum information from entanglement is an important and interesting quantity and has been studied not only in quantum gravity but 
also in condensed matter\cite{Amico} and quantum field theories. Suppose that a quantum system in isolation and thus in a pure state denoted by $\ket{\Psi}\in \mathcal{H}$, or density matrix $\rho=\ket{\Psi}\bra{\Psi}$ consists of a subsystem $A$ and its environment $B$, correspondingly its Hilbert space is $\mathcal{H} =\mathcal{H}_A\otimes \mathcal{H}_B$. The reduced state on $A$ and $B$ are given by
\begin{equation}
\rho_A =\text{tr}_B\rho,\ \ \rho_B=\text{tr}_A\rho, 
\end{equation}
and the entanglement entropy is given by
\begin{equation}
S_A = -\text{tr}_A \rho_A \log \rho_A = -\text{tr}_B\rho_B\log\rho_B=S_B \label{EE}
\end{equation}
Other quantity that indicates the degree of entanglement between A and its environment B is Renyi entropy\cite{Renyi}, which is an extension of the entanglement entropy and is expressed by:
\begin{equation}
S_A^{(n)} = \frac{1}{1-n}\log \text{tr} \rho_A^n  \label{Renyi1}. 
\end{equation}

In the case that two subsystems $A_1$ and $A_2$ are interior of the universal system $U$: $U\supset A=A_1+A_2$,
there are two representative measures of entanglement between $A_1$ and $A_2$:logarithmic negativity (LN) and mutual information(MI)\cite{Eisert, Groisman}.  The LN is given by
\begin{equation}
\mathcal{E} = \log \text{tr}_A |\rho_A^{T_2}| \label{lognega},
\end{equation}
where $\rho_A^{T_2}$ is partial transposition with respect to $A_2$ and the MI is calculated from entanglement entropies $S_A,S_{A_1},S_{A_2}$ or Renyi entropies $S_A^{(n)},S_{A_1}^{(n)},S_{A_2}^{(n)}$. 

After the seminal work by Hawking and Bekenstein\cite{Hawking,Bekenstein}, which showed that the entropy of a black hole is proportional to the area of its event horizon, the so-called holographic principle was proposed which asserted that quantum theoretical processes in $d_U$ dimensional space could be explained by processes in one smaller dimension,i.e.,the surface area\cite{tHooft,Maldacena}. Furthermore, in the two dimensional space($d_U=2$), when the boundary between two adjacent subsystems contains sharp corners of opening angles $\theta_k,k=1,2,\ldots$, it is said that the many quantities related to the entanglement can be expanded as 
\begin{equation}
 \alpha_B P_B/\delta-\sum_k \beta(\theta_k) \log P_B/\delta +\cdots, \label{expand1}
\end{equation}
where $P_B$ is the the boundary length between $A$ and $U$ or between $A_1$ and $A_2$,  $\delta$ is the UV cutoff\cite{Bueno}. The coefficient $\alpha_B$ correspondent to the area law depends on the cut-off $\delta$. On the other hand, the corner functions $\beta(\theta_k)$s are universal, independent of the way the continuous limit to the underlying theory is taken. In three dimensional space $d_U=3$, when the boundary surface with area $P_B$ has edges with length $E_j$ and dihedral angles $\theta_j,(j=1,2,\ldots)$ as well as sharp vertices with solid angles $\Omega_k$, the expansion is given by 
\begin{equation}
\alpha_B P_B/\delta^2+\sum_k \beta(\Omega_k) \log P_B/\delta^2 - \sum_j e(\theta_j)E_j/\delta  +  \cdots,  
\label{expand2}
\end{equation}
where the coner functions $\beta(\Omega_k)$s are universal similar to the those of formula(\ref{expand1}), while the edge term coefficients $e(\theta_j)$s are not\cite{Devakul}. 
 
For a two dimensional(2D) harmonic lattice with periodic boundary conditions (PBCs), which is a discretized massless scalar field in a torus-like space, the Ref\cite{Nobili} shows that the values of the corner functions of $\theta=\pi/2,\pi/4,3\pi/4$ obtained from LNs are highly consistent with those of conformal field theory describing the scalar field. It is also claimed that the values of the corner functions obtained from MIs corresponding to the Renyi entropies of order 1/2 are consistent with the above LN-derived values. In the computing process, we take the relation between the scale $L$ of the universal set $U$, the scale $l$ of $A$ and the lattice spacing $a$ to be $L\gg l\gg a$, so that the calculation on the lattice can approximate the based field theory which is the continuous limit. With those scale conditions, even when the canonical variables on the lattice sites of $U$ satisfy the fixed boundary conditions(FBCs), the values of those corner function in the region sufficiently far from the boundary of $U$ are expected to be consistent with the above ones calculated under the PBCs. 

In this paper, according to Ref.\cite{Nobili}, we first consider, as $A_2$, a set of lattice sites in several figures with corners of open angles $\theta=\pi/2, \pi/4, 3\pi/4$ on their boundaries and scales defined by the number of sites on the shortest side ranging from six to a maximum of about thirty. We suppose that $A$ is a set of sites in a square whose scale is 3 to 4 times larger than that of $A_2$ and sites of the set $A_1=A-A_2$ encloses $A_2$ sites. We calculate LNs and MIs between $A_1$ and $A_2$ for each figure while changing the scales of $A_2$ and A.  We fit those quantities by the formula (\ref{expand1}) to obtain the corner functions and their standard deviations in order to verify those values in the reference. Next, for the case where the canonical variables on the lattice sites satisfy the FBCs, we perform the same procedure as before at many locations in $U$ and observe whether the values of each corner function near the center of $U$ are consistent with those under the PBCs and how they change approaching the boundary of $U$.  

For a three dimensional lattice, in the same way as above, we compute the Renyi entropy of order $1/2$ for subsystems with the shape of cube and triangular prism whose base is an isosceles right triangle varying their scales. While using the expansion formula (\ref{expand2}) and confirming the existence of the edge term, we obtain the corner functions and edge term coefficients, respectively, of solid angles $\Omega=\pi/2$,$\pi/4$ and the dihedral angles $\theta=\pi/2$,$\pi/4$.

This paper is organized as follows: In section 2, we take the Hamiltonian of harmonic lattice which is the discretized massless scalar field. We present analytical solutions $p_{\vec{r}}$ and $q_{\vec{r}}$ of the Hamilton equations under PBCs and FBCs, as well as their vacuum correlation functions $X=\braket{0|q_{\vec{r}},q_{\vec{s}}|0}$ and $P=\braket{0|p_{\vec{r}},p_{\vec{s}}|0}$, where $\vec{r}$ and $\vec{s}$ are position vectors of lattice sites. Given the specific shapes and scales of the universal set $U$ and subsets $A$, $A_1$, $A_2$, after reviewing the prescription to compute LN and  various entropies from $X$ and $P$ matrices, we obtain the value of each corner function of $\theta=\pi/2, \pi/4, 3\pi/4$ (and each standard deviation) for harmonic lattice with PBCs and examine whether they are consistent with the previous study. 
In section 3, we compute the corner functions at many locations in $U$ which is a set of lattice sites under FBCs by the same computational prescription in Section 2, and analyze them. 
In section 4, we examine, in a 3D lattice, the corner functions and the edge term coefficients, respectively, of solid angles and of dihedral angles $\pi/2$,$\pi/4$ of Renyi entropy of order $1/2$. In section 5, we present some conclusions. 
\section{Models of harmonic lattices and entanglement indices computation}
We consider a free scalar field theory in $(d_U+1)$ dimensional spacetime of which Hamiltonian is 
\begin{equation}
H = \int d^{d_U} x\left[\frac{1}{2}(\pi(x)^2+(\nabla\phi(x))^2)+\frac{1}{2}m^2(\phi(x))^2\right],
\end{equation} 
and discretize it into a rectangle lattice consisting of  harmonic oscillators coupled with neighbor sites by spring-like interaction in the $d_U$-dimensional Euclidean space. Through a suitable normalization, $\pi(x)\to p_{\vec{j}}$, $\phi(x) \to q_{\vec{j}}$, $\int d^{d_U}x \to \sum_{\vec{j}}$ and the distance between two adjacent sites $a=1$, the Hamiltonian becomes to 
\begin{equation}
H = \frac{1}{2}\sum_{\vec{j}}(p_{\vec{j}}^2+\omega^2q_{\vec{j}}^2 +\sum_{\alpha =1}^{d_U}(q_{\vec{j}+\vec{e}_\alpha}-q_{\vec{j}})^2), \label{Hamil2}
\end{equation}
where $\vec{j}=(j_1,j_2,\dots,j_l,\ldots,j_{d_U})$ denotes the position of a site and $\vec{e}_\alpha=(e_1=0,0,\dots,e_\alpha=1,0,\ldots,0)$ is a unit vector on a spatial axis. The Hamiltonian equations are given by
\begin{align}
& \dot{q}_{\vec{j}} = p_{\vec{j}} \\
& \dot{p}_{\vec{j}} = \ddot{q}_{\vec{j}}= -\omega^2q_{\vec{j}} 
-\sum_{\alpha}\left((q_{\vec{j}} -q_{\vec{j}+\vec{e}_\alpha}) +(q_{\vec{j}}-q_{\vec{j}-\vec{e}_\alpha})\right) \nonumber \\
& = -\omega^2q_{\vec{j}} -\sum_\alpha(2 q_{\vec{j}} -q_{\vec{j}+\vec{e}_\alpha}-q_{\vec{j}-\vec{e}_\alpha}).
\end{align} 
Let us figure out solutions of the equations(Eqs). When we assume PBCs along directions of spatial axes, 
$
q_{\vec{j}+L_{\alpha} \vec{e}_{\alpha}}=q_{\vec{j}}, \ p_{\vec{j}+L_\alpha \vec{e}_{\alpha}}=p_{\vec{j}}
,\ \alpha=1,2,\ldots, d_U
$, the space of the universal system is a $d_U$-dimensional torus. Putting $V = \displaystyle{\prod_\alpha L_\alpha}$, 
solutions are obtained by 
\begin{align}
q_{\vec{j}} & = \sum_{\vec{k}}\frac{1}{\sqrt{2\omega_{\vec{k}}^p V}}\left[a_{\vec{k}}\exp(-i\omega_{\vec{k}}^p t+\sum_\alpha \frac{2\pi i}{L_\alpha}j_\alpha k_\alpha)
+a_{\vec{k}}^\dagger \exp(i\omega_{\vec{k}}^p t - \sum_\alpha \frac{2\pi i}{L_\alpha}{j_\alpha k_\alpha})\right], \\
p_{\vec{j}} & = -i\sum_{\vec{k}}\sqrt{\frac{\omega_{\vec{k}}^p}{2V}}\left[a_{\vec{k}}\exp(-i\omega_{\vec{k}}^p t+\sum_\alpha \frac{2\pi i}{L_\alpha}j_lk_\alpha)
-a_{\vec{k}}^\dagger \exp(i\omega_{\vec{k}}^p t - \sum_\alpha \frac{2\pi i}{L_\alpha}{j_\alpha k_\alpha})\right],
\end{align}
where $k_\alpha $'s are integers such that $0\leq k_\alpha \leq L_\alpha-1$. With these solutions, we get the dispersion relations
\begin{equation}
\omega_{\vec{k}}^p = \sqrt{\omega^2 + 4\sum_\alpha \sin^2(\pi k_\alpha/L_\alpha)}\ , \label{omegakp}
\end{equation}
and correlation functions of generalized coordinate $q_{\vec{r}}$ and their conjugate momentum $p_{\vec{r}}$ expressed as follows\cite{Nobili}:
\begin{align}
X_{\vec{r}\vec{s}}=\braket{q_{\vec{r}}q_{\vec{s}}} &  = \frac{1}{2V}\sum_{\vec{k}} \frac{1}{\omega_{\vec{k}}^p}\prod_\alpha \cos [2\pi k_\alpha(r_\alpha -s_\alpha )/L_\alpha],
\label{corpq}
\\
P_{\vec{r}\vec{s}}=\braket{p_{\vec{r}}p_{\vec{s}}} & = \frac{1}{2V}\sum_{\vec{k}}\omega_{\vec{k}}^p\prod_\alpha \cos [2\pi k_\alpha(r_\alpha-s_\alpha)/L_\alpha]. \label{corpp}
\end{align}
In these Eqs, $\omega$ is a small positive number which plays the roll of infrared cut-off. Furthermore, when performing the numerical calculations of these matrices specially in $X$ matrix of Eq.(\ref{corpq}), stable  values can be obtained by taking the continuous limit of $k_l$s and replacing the sum over them with integrals, and then actually summing over the half-integer of $k_l$s.

On the other hand, when we take FBCs, 
\begin{align*}
q_{(j_1,j_2,\ldots, j_\alpha=L_{\alpha},\ldots,j_{d_U})}  & =q_{(j_1,j_2,\ldots,\ j_{\alpha}=0,\ \ldots,j_{d_U})}=0 , \\
\ p_{(j_1,j_2,\ldots, j_\alpha=L_{\alpha},\ldots,j_{d_U})} & =p_{(j_1,j_2,\ldots,\ j_\alpha=0\ ,\ldots,j_{d_U})}=0, \\
&  \alpha=1,2,\ldots, d_U,
\end{align*}
the universal system is a $d_U$-dimensional rectangular solid and putting $V = \prod_\alpha L_\alpha$ as well as the case of the PBCs,  we have found the solutions 
as follows:
\begin{align}
q_{\vec{j}} & = \sum_{\vec{k}}\sqrt{\frac{1}{2\omega_{\vec{k}}^f V}}
\left(a_{\vec{k}}\exp(-i\omega_{\vec{k}}^f t)
+a_{\vec{k}}^\dagger \exp(i\omega_{\vec{k}}^f t)\right)\prod_\alpha \sqrt{2}\sin\left(\frac{\pi }{L_\alpha}j_\alpha k_\alpha \right), \\
p_{\vec{j}} & = -i\sum_{\vec{k}}\sqrt{\frac{\omega_{\vec{k}}^f}{2V}}
\left(a_{\vec{k}}\exp(-i\omega_{\vec{k}}^f t)
-a_{\vec{k}}^\dagger \exp(i\omega_{\vec{k}}^f t)\right)\prod_\alpha \sqrt{2}\sin\left(\frac{\pi}{L_\alpha}j_\alpha k_\alpha \right), 
\end{align} 
where $\omega_{\vec{k}}^f$ is given by
\begin{equation}
\omega_{\vec{k}}^f  = \sqrt{\omega^2 + 4\sum_\alpha \sin^2(\pi k_\alpha/(2L_\alpha))}, \label{omegakf}
\end{equation}
and correlators are given by
\begin{align}
X_{\vec{r}\vec{s}}=\braket{q_{\vec{r}}q_{\vec{s}}} 
& = \frac{1}{2V}\sum_{\vec{k}}\frac{1}{\omega_{\vec{k}}^f}
\prod_\alpha 2\sin\left(\frac{\pi r_\alpha k_\alpha}{L_\alpha}\right)\sin\left(\frac{\pi s_\alpha k_\alpha}{L_\alpha}\right), \label{corfq}\\
P_{\vec{r}\vec{s}}=\braket{p_{\vec{r}}p_{\vec{s}}} 
& = \frac{1}{2V}\sum_{\vec{k}}\omega_{\vec{k}}^f
\prod_\alpha 2\sin\left(\frac{\pi r_\alpha k_\alpha}{L_\alpha}\right)\sin\left(\frac{\pi s_\alpha k_\alpha}{L_\alpha}\right).\label{corfp}
\end{align}
In the FBCs case, for any spatial axis direction, $j_\alpha$ ranges 0 to $L_\alpha$ and the number of $j_\alpha $ is $L_\alpha+1$, the dynamical variables $q_{(j_1,j_2,\ldots,j_\alpha,\ldots,j_{d_U})}$, $p_{(j_1,j_2,\ldots,j_\alpha,\ldots,j_{d_U})}$ exist only at $1\leq j_\alpha \leq L_\alpha-1$ and $k_\alpha$s also take  integers in the range $1\leq k_\alpha\leq L_\alpha-1$. Numerical computations of these $X$,$P$ matrices yields stable values, even when summed over an integer of $k_\alpha$s.

The Hamiltonian (\ref{Hamil2}) can be rewritten as follows:
\begin{equation}
H=\frac{1}{2}\sum_{\vec{r}} p_{\vec{r}}^2+\frac{1}{2}\sum_{\vec{r}}\sum_{\vec{s}}q_{\vec{r}}K_{\vec{r}\vec{s}}q_{\vec{s}}. 
\label{HalK} 
\end{equation}
In the FBCs case, the sites $\vec{r}$ and $\vec{s}$ are in the range $1\leq r_\alpha,s_\alpha\leq L_\alpha-1$, the sites where dynamical variables exist. The nonzero components of the matrix $K$ are given by
\begin{equation}
 K_{\vec{r}\vec{s}} 
 =  \begin{cases}
      2d_U+\omega^2 & \text{if}\ \vec{r}=\vec{s} \\ 
     -1 & \text{if $\vec{r}$ and $\vec{s}$ are neighboring sites}
\end{cases}. \label{Krs}
\end{equation} 
In the PBCs case, the sites $\vec{r}$ and $\vec{s}$ are in the basic area $0 \leq r_\alpha ,s_\alpha \leq L_\alpha -1$. If $\vec{r}$ and $\vec{s}$ are points relative to each other on opposite $d_U$-rectangular solid faces, 
$\vec{r}=(r_1,\ldots,r_{\alpha-1},0,r_{\alpha+1}\ldots,r_{d_U})$, $\vec{s}=(s_1,\ldots,s_{\alpha-1},L_{\alpha}-1,s_{\alpha+1},\ldots,s_{d_U})$, $K_{\vec{r}\vec{s}}$ and $K_{\vec{s}\vec{r}}$ are $-1$ in addition to (\ref{Krs}). The diagonalizations of those Hamiltonians provide the values $(\omega_{\vec{k}}^p)^2$ of (\ref{omegakp}) and  $(\omega_{\vec{k}}^f)^2$ of (\ref{omegakf}) as the eigenvalues of $K$ matrices.  

For the ground state of (\ref{HalK}), the values of correlators (\ref{corpq}),(\ref{corpp}), (\ref{corfq}),(\ref{corfp}) are also to be obtained from the matrix $K$ as follows \cite{Lykken,Yazdi}:
\begin{align}
X_{\vec{r}\vec{s}} & =\braket{q_{\vec{r}}q_{\vec{s}}} = \frac{1}{2}(K^{-1/2})_{\vec{r}\vec{s}}, \label{corX}\\
P_{\vec{r}\vec{s}} & =\braket{p_{\vec{r}}p_{\vec{s}}} = \frac{1}{2}(K^{1/2})_{\vec{r}\vec{s}}.  \label{corP}
\end{align}

Consider a set of sites inside the universal system $U$ described in the previous section, and denote it as a subsystem $A$. Also, let its environment $B=U-A$. Using the correlators (\ref{corpq}),(\ref{corpp}), (\ref{corfq}),(\ref{corfp}), or (\ref{corX}),(\ref{corP}), we can obtain the Renyi entropies $S_A^{(n)} $and entanglement entropy $S(A)$\cite{Bombelli,Nez}. First, we make the matrices of rows and columns of the above correlators restricted to those corresponding to the sites of the subsystem, i.e., $X_{\vec{u}\vec{v}}$ and $P_{\vec{u}\vec{v}}$, where $\vec{u},\vec{v}$ are the positions of sites in the subsystem $A$. Second, we compute the matrix $\Xi =\sqrt{XP}$ and its eigenvalues $\xi_i$.
Using the eigenvalues, we can describe the entanglement entropy(\ref{EE}) and Renyi entropies(\ref{Renyi1}) as follows:
\begin{align}
S_A &  
= \sum_{i=1}^{\dim(\mathcal{H}_A)}\left(
(\xi_i+\frac{1}{2})\log(\xi_i+\frac{1}{2})-(\xi_i-\frac{1}{2})\log(\xi_i-\frac{1}{2})
\right). \label{EEbyC} \\
S_A^{(n)} &  
= \frac{1}{n-1}\sum_{i=1}^{\dim(\mathcal{H}_A)}
\log\left( (\xi_i+\frac{1}{2})^n - (\xi_i-\frac{1}{2})^n\right) \label{Renyi2}
\end{align} 

Furthermore, we divide the subsystem $A$ into subsubsystems $A_1$ and $A_2$, $ A=A_1+A_2$. Let us consider mutual information (MI) and logarithmic negativity (LN) as indices of the entanglement between $A_1$ and $A_2$\cite{Groisman,Nobili}. The MI and corresponding combination of Renyi entropies are given by
\begin{equation}
I_{A_1,A_2}=S_{A_1}+S_{A_2}-S_A=\lim_{n \to 1}I_{A_1,A_2}^{(n)},
\end{equation}
and $I_{A_1,A_2}^{(n)} =S_{A_1}^{(n)}+S_{A_2}^{(n)}-S_A^{(n)}$.

From the matrices $X,P$, we can also calculate LN  between the subsystems. First, we find the partial transposition of the density matrix $\rho_A$  with respect to $A_2$, which corresponds to the time reversal only to the degrees of freedom in $A_2$\cite{Simon}. Therefore, it does not change the matrix $X$. On the otherhand, as for the matrix $P$, when we rewrite the matrix into the form as:
\begin{equation}
P = \begin{pmatrix}
P_{u_1 u_1} & P_{u_1 u_2}\\
P_{u_2 u_1} & P_{u_2 u_2}
\end{pmatrix},
\end{equation}
where $P_{u_i u_j}$ is the submatrices restricted to sites with rows inside $A_i$ and columns inside $A_j$, the partial transposition can be written as follows:
\begin{equation}
P^{T_2} = 
\begin{pmatrix}
P_{u_1 u_1} & -P_{u_1 u_2}\\
-P_{u_2 u_1} & P_{u_2 u_2}
\end{pmatrix}.
\end{equation}
Second, we compute the matrix $\Xi' =\sqrt{XP^{T_2}}$ and its eigenvalues $\xi'_i$, which leads to the LN:
\begin{equation}
\mathcal{E}=\sum_{i=1}^{\dim(\mathcal{H}_A)}\log[\max(1,(2\xi'_i)^{-1})]
\end{equation}

The MI and LN between adjacent sets $A_1$ and $A_2$ are given by the length $P_B$ of the shared boundary $SB=\partial A_1 \cap \partial A_2$, as follows:
\begin{align}
I_{A_1,A_2}^{(n)} & =2 \tilde{a} P_B -\tilde{b}_{\text{total}}\cdot\log P_B + \tilde{c}_0+\tilde{c}_{-1}P_B^{-1}+\tilde{c}_{-2}P_B^{-2}+\cdots ,\nonumber\\
& \tilde{b}_{\text{total}}  = \sum_{\text{vertices of}\ SB }(\tilde{b}(\theta_i^{(1)})+\tilde{b}(\theta_i^{(2)}))
  \label{MIandbtotal} \\
\mathcal{E} & = a P_B -b_{\text{total}}\cdot\log P_B + c_0+c_{-1}P_B^{-1}+c_{-2}P_B^{-2}+\cdots   \nonumber\\
 & b_{\text{total}}  = \sum_{\text{vertices of}\ SB }b(\theta_i^{(2)}) \label{LNandbtotal}
\end{align}
where $\tilde{b}(\theta_i^{(j)})$ and $b(\theta_i^{(j)})$ are so called corner functions and $\theta^{(j)}_i$ is the angle corresponding to the $i-$th vertex of $SB$ belongs to $A_j$\cite{Nobili}. 

As a starting point for numerical calculations of these corner functions, we consider as the universal system $U$ a set of sites on a 2D lattice in a square with 1200 sites on a side in the x- and y-directions satisfying PBCs, so that $L_1, L_2=1200$. As the region for computing the correlation function values (\ref{corpq}) and (\ref{corpp}), we take a square region $D$ with 110 sites on each side parallel to those of $U$. In general, these computations take an enormous amount of time in the case that the $U$ and $D$ scales are large, but using Yukawa-21 at Yukawa Institute for Theoretical Physics in Kyoto University, we have significantly shortened the time. First, as the region $D$, we take a $110\times 110$ square region with the same centroid as $U$.  Only in cases that the subsystem $A$ is in the region $D$, we shall calculate the MIs and LNs between subsystems $A_1$ and $A_2$. According to \cite{Nobili}, we take, as the subsystem $A_2$, the set of sites within a square, a right isosceles triangle and a trapezoid, denoted by $A_s$, $A_r$, $A_t$ with vertex angles $\pi/2$, $\pi/4$, $3\pi/4$. Similarly, a parallelogram with the vertex angles $\pi/4,\ 3\pi/4$ is also considered and denoted by $A_p$. And we assume that these sites are enclosed by sites of the subsystem $A_1$.

\begin{figure}[H]
\includegraphics[height=68mm, bb = 0 0 1083 316]{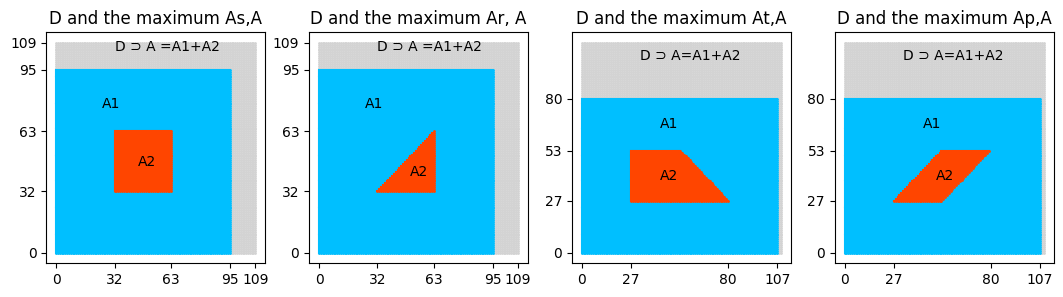}
\caption{$D$ and $A_2$, $A$ of the maximum scale: The vertical and horizontal scales represent the coordinates of the sites corresponding to the vertices of the regions $A$ and $A_2$ with the site at the lower left corner of the region $D$ as the origin. }
\label{DAA2}
\end{figure}

We define the scale $l$ and its perimeter $P_B$ of each shape as follows:
For $A_s$, letting the number of sites on a side the scale $l$, the perimeter $P_B=4*l$. 
For $A_r$, letting the number of sites on one of the sides franking the right angle the scale $l$, $P_B=(2+\sqrt{2})(l+0.5)$. 
For $A_t$, letting the number of sites on the side corresponding to its height the scale $l$, $P_B=(4l-1)+\sqrt{2}l=(4+\sqrt{2})l-1$. 
For $A_p$, letting the number of sites on a side in the direction of the base the scale $l$, $P_B=2(1+\sqrt{2})l$. 
 For $A_s$ and $A_r$, we take as the system $A$ a square with the 3 times larger than respective scale, while for $A_t$ and $A_p$, a rectangle of 4 times scale in the direction of the base and 3 times scale in the direction of the height. The sides of the squares and rectangles are assumed to be parallel to those of the universal system. 
The MIs and LNs are computed by varying the scale of the $A_2$ system from $l_{\text{min}}=6$ to $l_{\text{max}}=32$ for $A_s$ and $A_r$, and from $l_{\text{min}}=6$ to $l_{\text{max}}=27$ for $A_t$ and $A_p$. Fig.\ref{DAA2} shows the largest $A_2$ systems and corresponding $A$ systems. 
\vspace{-2.6cm}\\   

We fit the values of MIs and LNs in the range of $P_B$ from the suitable minimum perimeter $P_{\text{min}}$ to the $P_{\text{end}}$ to Eqs (\ref{MIandbtotal}) and (\ref{LNandbtotal}) by the least squares method in order to obtain the constants $a,\tilde{a}$, $b_{\text{total}},\tilde{b}_\text{total}$, $c_{-k},\tilde{c}_{-k}$, $k=0,1,\ldots n$. We perform these operations changing $P_{\text{end}}$ up to $P_{\text{max}}$, the perimeter corresponding to the largest scale $l_{\text{max}}$ of the subsystem $A_2$ and obtain the mean values and standard deviations(SDs) of $\tilde{b}_{\text{total}}$, $b_{\text{total}}$. In the range from 0 to 4 of $n$,  the smallest SD of $\tilde{b}_\text{total}$s of (\ref{MIandbtotal}) which determine how far $k$ of which parameters $\tilde{c}_{-k}$ are to be fitted, for all shapes of $A_2$ are obtained for $n=1$. On the other hand, those of $b_\text{total}$s of (\ref{LNandbtotal}) are $n=2$ for $A_s$ and $A_r$,  $n=1$ for $A_t$ and $A_p$.  Hereafter, the values of $\tilde{b}_\text{total},b_\text{total}$  and their SDs are denoted by $\tilde{b}_s,b_s$, $\tilde{b}_{ss},b_{ss}$, $\tilde{b}_r,b_r$, $\tilde{b}_{rs},b_{rs}$, etc., respectively, depending on the shape of $A_2$. 

Let us discuss the corner function values obtained from LNs, which, as will be discussed later, have smaller SDs, i.e., have more stable values than those obtained from MIs. Corner functions $b(\pi/2)$, $b(\pi/4)$ and their SDs are given as follows.
\begin{align}
\begin{split}
b(\pi/2)=\ & b_s/4,\ \ \ \ \ \ \ \ \ \ \ \ \ \ \ \ \ \ b(\pi/2)_{\text{sd}}=b_{ss}/4, \\
b(\pi/4)=\ & b_r/2-b_s/8,\ \ \ \ \ \ \ \ \ b(\pi/4)_{\text{sd}}=\sqrt{(b_{rs}/2)^2+(b_{ss}/8)^2}, 
\end{split}
\label{bpi_2pi_4}
\end{align}
while $b(3\pi/4)$ and $b(3\pi/4)_{\text{sd}}$ can be given in two ways using $b_t$, $b_{ts}$ or $b_p$,$b_{ps}$:
\begin{align}
 \begin{split}
  b(3\pi/4)=\ & b_t-b_r/2-3b_s/8,\ \ b(3\pi/4)_{\text{sd}}=\sqrt{b_{ts}^2+(b_{rs}/2)^2+(3b_{ss}/8)^2},\\
  b(3\pi/4)=\ & b_p/2-b_r/2+b_s/8,\ \ \ \ b(3\pi/4)_{\text{sd}}=\sqrt{(b_{ps}/2)^2+(b_{rs}/2)^2+(b_{ss}/8)^2},
 \end{split}
\label{b3pi_4}
\end{align}
The numerical results of the Eqs (\ref{bpi_2pi_4})  are as follows:
\begin{align}
 \begin{split}
  b(\pi/2)=0.02954,& \ \ \ b(\pi/2)_{\text{sd}}=0.00013, \\
  b(\pi/4)=0.09670,& \ \ \ b(\pi/4)_{\text{sd}}=0.00032, 
 \end{split} \label{WPbpi_24}
\end{align}
while the values of (\ref{b3pi_4}) are respectively given by 
\begin{align}
 \begin{split}
  b(3\pi/4)=0.00604,& \ \ \ \ b(3\pi/4)_{\text{sd}} =0.00047, \\
  b(3\pi/4)=0.00635,& \ \ \ \ b(3\pi/4)_{\text{sd}} =0.00038. 
 \end{split} \label{WPb3pi_4}
\end{align}
These corner function values are almost equal to those in the Ref.\cite{Nobili}, and are stable as the relative standard deviations are less than 0.004 or less than 0.07. However, in detail, the difference of $b(3\pi/4)$ from the literature is less than its SD, while those of $b(\pi/2)$ and $b(\pi/4)$ are about 4 and 3 times larger than their SDs, respectively.

Let us now discuss the corner functions obtained from Renyi entropy-based MIs. In particular, the corner functions $\tilde{b}(\theta)$ due to Renyi entropy with the order $1/2$ should coincide with $b(\theta)$ due to LNs so that $b(\theta)$ and $\tilde{b}(\theta)$ are reasonable checks for each other \cite{Nobili}. Since for those corner functions $\tilde{b}(2\pi-\theta)=\tilde{b}(\theta)$, $0\leq \theta\leq \pi$ and $\tilde{b}_{\text{total}}$ of the Eq.(\ref{MIandbtotal}) are rewritten to 
$$ 
\tilde{b}_{total}= \sum_{\text{vertices of SB}} 2\tilde{b}(\theta_i^{(2)}). 
$$
Then, the relationship between the values of each $\tilde{b}(\theta),\theta=\pi/2,\pi/4,3\pi/4$ and $\tilde{b}_{\text{total}}$s are obtained by not only changing each '$b$' to '$\tilde{b}$' but also replacing each $b_{\text{total}}$ to $\tilde{b}_{\text{total}}/2$ in equations (\ref{bpi_2pi_4}) and (\ref{b3pi_4}): $\tilde{b}(\pi/2)=\tilde{b}_s/8$, $\tilde{b}(\pi/2)_{\text{sd}}=\tilde{b}_{ss}/8$, etc..  The corner function values obtained from MI in the above manner are as follows:
\begin{align}
 \begin{split}
  \tilde{b}(\pi/2)=0.02962,& \ \ \ \tilde{b}(\pi/2)_{\text{sd}}=0.00003, \\
  \tilde{b}(\pi/4)=0.1004,& \ \ \ \tilde{b}(\pi/4)_{\text{sd}}= 0.0011, 
 \end{split} \label{WPtbpi_24}
\end{align}
and 
\begin{align}
 \begin{split}
  \tilde{b}(3\pi/4)=0.0052,& \ \ \ \ b(3\pi/4)_{\text{sd}} =0.0012,\\
  \tilde{b}(3\pi/4)=0.0038,& \ \ \ \ b(3\pi/4)_{\text{sd}} =0.0013. 
 \end{split} \label{WPtb3pi_4}
\end{align}
The values of these corner functions are equal to the values obtained from LNs, equations (\ref{WPbpi_24}) and (\ref{WPb3pi_4}), within a range of a few times the respective SDs. However, the ratios of $\tilde{b}(3\pi/4)$ to the values obtained from LN are 0.86 or 0.6, which is not a sufficient agreement, and as the relative standard deviations are as large as 0.23 or 0.34, they are not stable values. 

At the end of the calculation for a lattice satisfying PBCs, we take a square region $D$ whose centroid is $(99.5,99.5)$ with the site at the lower left corner of $U$ as the origin and repeat the same calculations as above. The results showed that for all corner functions treated, the values and their SDs are in agreement with (\ref{WPbpi_24}), (\ref{WPb3pi_4}), (\ref{WPtbpi_24}) and (\ref{WPtb3pi_4}) up to the most significant digit of each SD that are to be expected with the universal system $U$ under PBCs.
\section{Harmonic lattice with fixed boundary conditions}
\begin{figure}[H]
\hspace{4cm}
\includegraphics[height=68mm, bb = 0 0 1083 316]{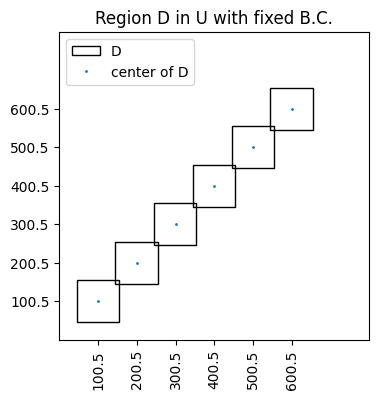}
\caption{Locations of the regions $D$ in which correlators are computed within the universal system $U$ of FBCs. \label{DinWF} 
}
\end{figure}

\begin{figure}[H]
    \begin{tabular}{cc}
        \begin{minipage}{.5\textwidth}
            \centering
            \includegraphics[width=0.9\linewidth]{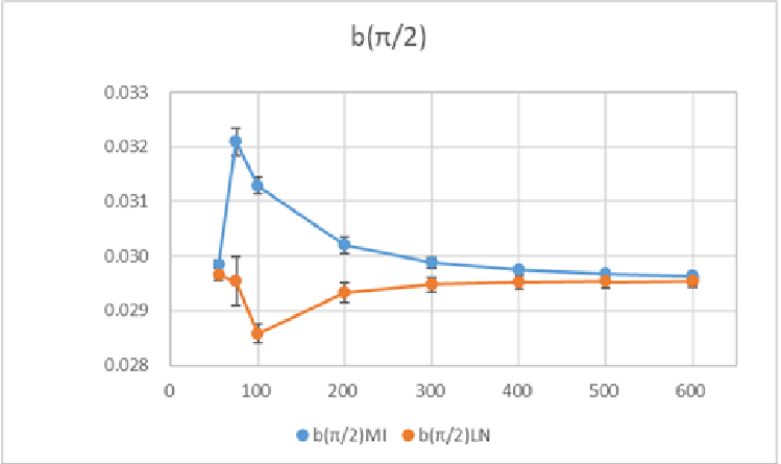}
            \caption*{ } 
        \end{minipage}
        \begin{minipage}{.5\textwidth}
            \centering
            \includegraphics[width=0.9\linewidth]{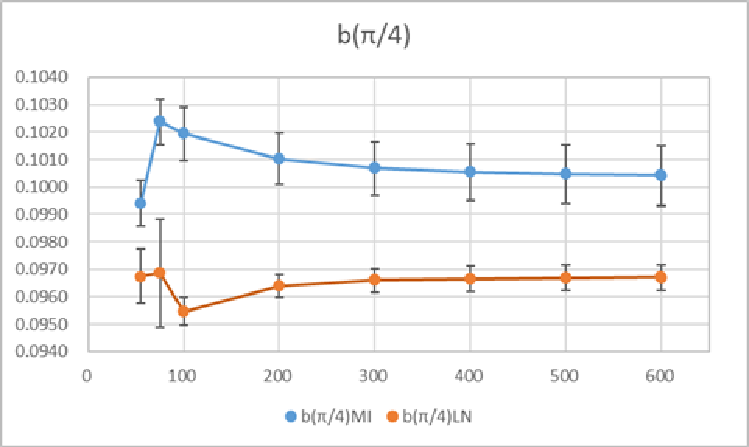}
            \caption*{ } 
         \end{minipage} \\
        \begin{minipage}{.5\textwidth}
            \centering
            \includegraphics[width=0.9\linewidth]{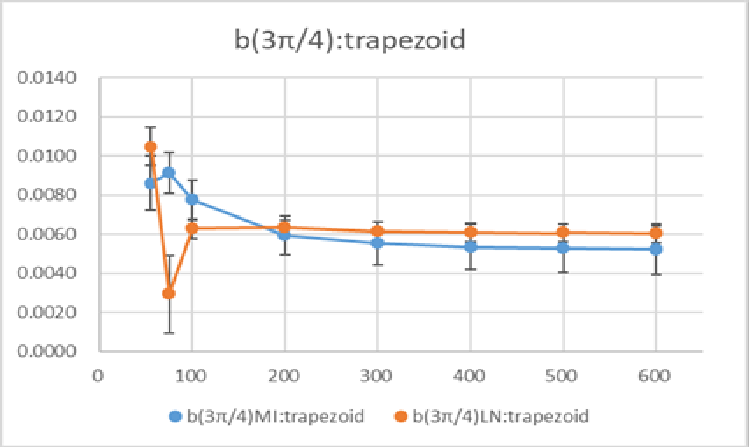}
            \caption*{ } 
        \end{minipage}
        \begin{minipage}{.5\textwidth}
            \centering
            \includegraphics[width=0.9\linewidth]{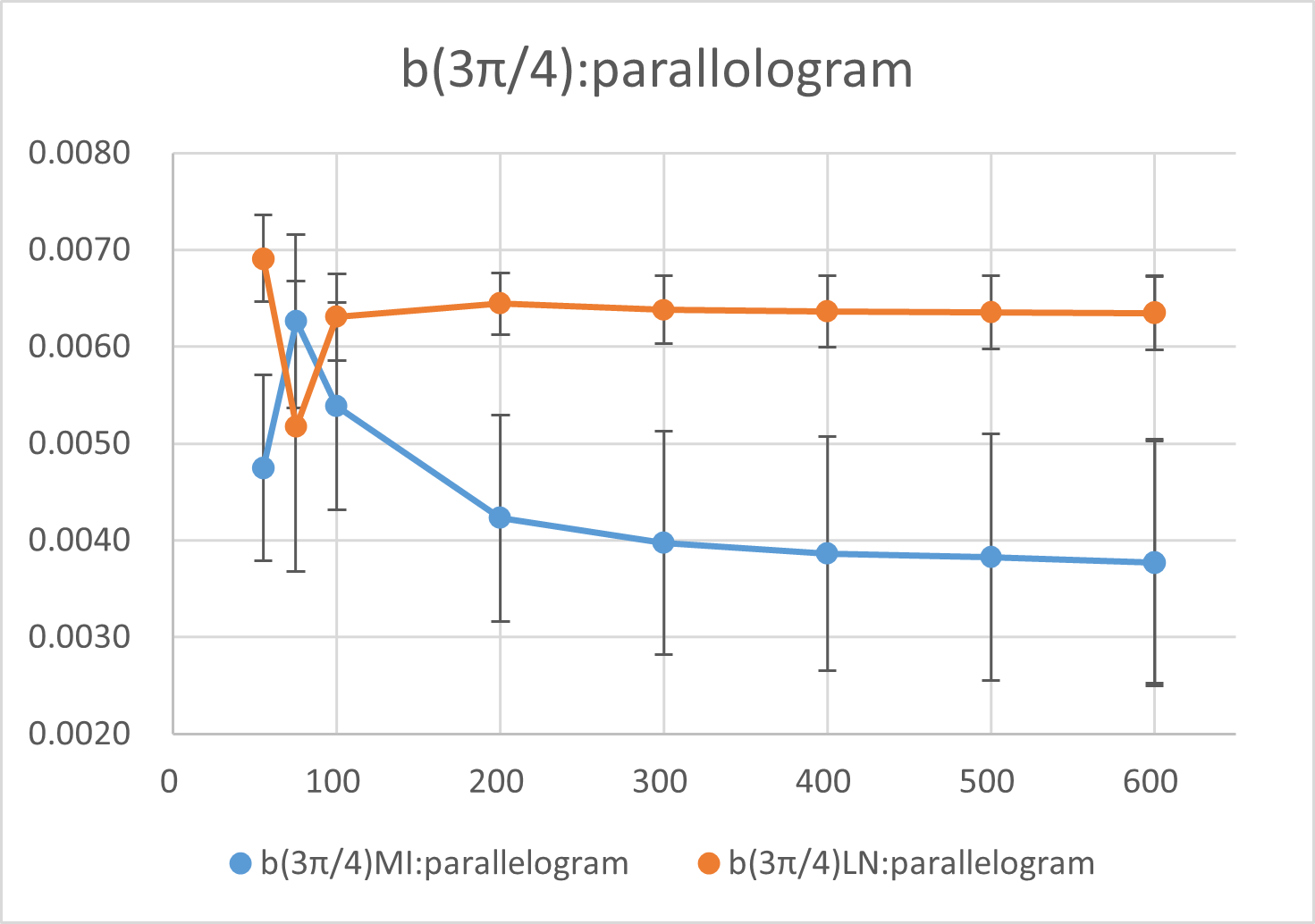}
            \caption*{ } 
         \end{minipage}
    \end{tabular}
\vspace{-0.9cm}\\
\caption{Values of corner functions obtained in various $D$ regions. A scale value of the horizontal axis of each graph indicates (the $x$-coordinate of the centroid of $D$ region)$-0.5$. The graphs in the lower row are values of $b(3\pi/4)$ obtained using $A_t$ and $A_p$, respectively.} 
\label{graph_bs}
\end{figure}

In this section, as an universal system $U$, we consider a set of sites in a square satisfying FBCs  of the same scale as in the previous section. Although the square has 1202 sites on a side, due to the FBCs for the two sites at each end, the number of sites with dynamic variables is 1200 and $L_1,L_2=1201$. Also, for the subsystems $A_1$, $A_2$, $A=A_1+A_2$, we have the same shapes and scales as in the previous section (for $A_2$, square $A_s$, right isosceles triangle $A_r$, trapezoid $A_t$, parallelogram $A_p$). On the other hand, the region $D$, which includes $A$ and for which the correlation functions of Eqs.(\ref{corfq}) and (\ref{corfp}) are to be calculated is taken at several locations from the centroid of $U$ to near the lower left corner of $U$. We adopt the six positions of $D$ in Fig.\ref{DinWF} as well as two regions  whose centroid are (55.5,55.5) and (75.5,75.5) with the origin at the fixed end site at the lower left corner of $U$.  The last two positions of $D$ are added to observe the fixed end effect to corner functions. For each of them, we compute $b_{\text{total}}$s and $\tilde{b}_{\text{total}}$s in Eqs.(\ref{MIandbtotal}),(\ref{LNandbtotal}) with figures of $A_2$ and corner functions involved as in the previous section. 
\vspace{-1cm}\\   
In the LN computations, for four $D$ regions closer to the centroid of $U$, the SDs are minimized when fitting to equation (\ref{LNandbtotal}) up to $c_{-2}$ for $A_s$ and $A_r$ and up to $c_{-1}$ for $A_t$ and $A_p$, as in the case in where $U$ satisfies the PBCs. For the sake of consistent computation, we adopt as the value of each $b_\text{total}$ the value obtained by the same fitting as above for the four cases where the regions of $D$ are closer to the lower left corner of $U$. On the otherhand in the MI computations, for two $D$ regions closer to the centroid of $U$, the SDs are minimized when fitting up to $c_{-1}$ for all shapes of subsystem $A_2$ and so, we adopt as $\tilde{b}_{\text{total}}$ the values fitting up to $c_{-1}$ for all shapes of the other six $D$ regions, too. 
Each graph in Fig.\ref{graph_bs} plots the variation of each corner function corresponding to the location of the $D$ region with each SD as its error bar. 

The values of corner functions $\pm$ their SDs obtained from LNs calculated in the D region with the cenroid $(600.5,600.5)$ are as follows:
\begin{align}
 \begin{split}
 b(\pi/2)=0.02955\ \pm\ 0.00011,\\ 
 b(\pi/4)=0.09670\ \pm\ 0.00032, 
 \end{split}
\label{WFbpi_24}
\end{align}
and as for two values obtained from $A_t$ and $A_p$, 
\begin{equation}
  b(3\pi/4)=0.00604\pm 0.00047,\ \ b(3\pi/4)=0.00635\pm 0.00038. \label{WFb3pi_4}. 
\end{equation}
On the other hand, those values obtained from MIs are given by:
\begin{align}
 \begin{split}
  \tilde{b}(\pi/2)=0.02963 \pm 0.00003,\\ 
  \tilde{b}(\pi/4)=0.1004  \pm 0.0011, 
 \end{split}
\label{WFtb3pi_4}
\end{align}
and the two values of $\tilde{b}(3\pi/4)$ from $A_t$, $A_p$ are given by 
\begin{equation}
  \tilde{b}(3\pi/4)=0.0052 \pm 0.0013,\ \ \tilde{b}(3\pi/4)=0.0038 \pm 0.0013. \label{WFtb3pi_4}
\end{equation}
These values, respectively, are almost the same as those in Eqs.(\ref{WPbpi_24}) $\sim $ Eq.(\ref{WPtb3pi_4}) ( i.e., the values obtained in the universal system with PBCs and the $D$ region with the centroid (600.5,600.5). In each graph of Fig.\ref{graph_bs}, each corner function obtained in the universal system with PBCs is also plotted at the horizontal axis value of 600, but it is indistinguishable from the plotted values of Eq(\ref{WFbpi_24}) $\sim $ Eq(\ref{WFtb3pi_4}),  because they almost overlap. The results are expected, since, even when the universal system $U$ follows the FBCs, the effects of the fixed ends are expected to be small if the $D$ is sufficiently far from the boundary of the $U$. Furthermore, the values of each corner function within the horizontal axis coordinates (i.e., the center of the D region) between 300 and 600 are equal within their respective standard deviations, where the graph forms a horizontal line segment. Thus, for a sufficiently large system $U$, the effects of the fixed edge are to decrease in the region away from the edge of the $U$ about three to four times the size of the largest subsystem $A$, and the physical states of subsystems are expected to be the same as those of the harmonic lattice in the universal system obeying the PBCs. 

In the conformal theories(CFTs), i.e. , the continuous limit of the above lattice theories, each corner function $\tilde{b}(\theta)$ by Renyi entropy with the order $1/2$ has a universal lower limit, $\tilde{b}\geq \frac{1}{32\pi}(\pi-\theta)^2$\cite{Bueno}. Specifically,
$$
\tilde{b}(\pi/2)\geq 0.02454,\ \tilde{b}(\pi/4)\geq 0.05522,\ \tilde{b}(3\pi/4)\geq 0.006136.
$$
Looking at the plots with the horizontal axis coordinate 600 of each graph in Fig.3, the error bars, i.e., SDs of $b(\pi/2)$, $\tilde{b}(\pi/2)$ and $b(3\pi/4)$, $\tilde{b}(3\pi/4)$ calculated from trapezoidal $A_2$ in shape, $A_t$, overlap respectively, where the lower limits expected by the CFT are included. On the other hand, for corner functions of $\pi/4$ and those of $3\pi/4$ calculated from parallelograms, $A_p$, when the error bars are set to three or two times SDs, where includes the lower limit expected by the CFT, they overlap. 
\section{3D lattice systems: Edge contributions to entanglement }
So far, taking the universal system $U$ and the subsystems $A=A_1+A_2$ on a 2D lattice, 
we have analyzed the LNs and the MIs between $A_1$ and $A_2$. In this section, we investigate the entanglement between two adjacent sets of harmonic oscillators placed on lattice sites in $d_U=3$ dimensional space. For solid figures, the number of sites in the interior increases rapidly in proportion to the cube of the scale. Considering the limitation of available computer resources, we assume that the universal system $U$ is the set of sites in a cube with 120 sites on a edge. As a subsystem, we take the sets of sites near the center of gravity of the system $U$ that are inside a cube or a triangular prism whose base is a isosceles right triangle. Then, assuming $U=A+B$, we compute the Renyi entropy of order $1/2$ while varying the scale,the number of sites on the shortest edge of each shape of the subsystems $A$ from 6 to 20. In these cases, the boundary surfaces of the two subsystems $A$ and $B$ have not only sharp vertices but also edges. 
As for the contribution to entanglement from sharp corners and edges, the following equation is said to be valid from formula (\ref{expand2}) omitting cut-off\cite{Devakul,Sierens}:
\begin{equation}
S_A^{(1/2)} =\alpha P_B + \beta_{total} \log P_B -\sum_j e(\theta_j) E_j+   \cdots, \label{3DEE}
\end{equation}
where $\beta_{total}$ is the sum of corner function $\beta(\Omega_k)$ of sharp vertex with the solid angle $\Omega_k$, and $e(\theta_j)$ and $E_j$ are the edge term coefficient and the length of the edges with dihedral angle $\theta_j$. 
By fitting the above data of the Renyi entropy according to the formula (\ref{3DEE}), we obtain the corner functions $\beta(\pi/2)$, $\beta(\pi/4)$, the edge term coefficients $e(\pi/2)$ and $e(\pi/4)$ and their SDs. When doing so, as is done in equations (\ref{MIandbtotal}), (\ref{LNandbtotal}) and later, terms proportional to $P_B^{-k}(k=0,1,2...,n)$ are added to the right-hand side of the equation (\ref{3DEE}), and the operations are repeated while increasing $n$. Then, we select the $n$ that gives the most stable total values of corner functions, i.e., the smallest SD of the total value $\beta_{total}$. In fact, in both the PBC and FBC cases and for both $A$ shapes of cube and triangle prism, the most stable total corner functions are obtained by the fitting up to the constant term $c_0$.   

To begin, for the case where the shape of the subsystem $A$ is a cube, let us compare the sum of the corner functions and their SDs obtained by the method described above 
with those obtained by fitting to the Eq.(\ref{3DEE}) minus the edge terms. 
The table \ref{edge_corner1} gives the corner functions and their SDs when computing with and without edge terms for the case where $U$ satisfies the PBCs and FBCs, respectively.  In the case where the edge terms are considered, the coefficient values and their SDs are also included. 
\begin{table}[h]
\caption{Effect of the presence of an edge term on Renyi entropy in a 3-dimensional object}
\label{edge_corner1}
 \hspace{2.5cm}
 \begin{tabular}{|c|l|ll|}
 \hline
 & & $\beta(\pi/2)\pm \beta(\pi/2)_{sd}$ & $e(\pi/2) \pm e(\pi/2)_{sd}$ \\
 \hline 
 PBC & without $e$ & $0.242 \pm  0.026$ & \\
     & with $e$ & $0.005673 \pm 0.00010$ & $0.06725 \pm 0.00003$ \\
 \hline 
 FBC & without $e$ & $0.242 \pm 0.026$ & \\
    &  with $e$ & $0.005604 \pm 0.000059$ & $0.06723\ \pm 0.00002$ \\
 \hline
 \end{tabular}
\end{table}
Although the corresponding values in the PBC row and FBC row of the table are almost the same, the corner functions without considering the edge terms are about 43 times larger than those with the edge terms. Furthermore the relative SDs of corner functions are about 0.11 without considering the edge terms, while those are about 0.02 for PBCs and 0.01 for FBCs when the edge terms are considered.  In other words, when the edge terms are taken into account, the corner functions are five to ten times more stable than when the edge terms are not taken into account. For the edge term coefficients, very stable values 0.06725, 0.06723 with relative SDs of 0.0004, 0.0003 are obtained for the two lattice boundary conditions. The results indicate the existence of the edge terms in the Eq.(\ref{3DEE}). 

Next, let us discuss the case where the shape of $A$ system is a triangular prism. This triangular prism is with the base that is an isosceles right triangle formed by dividing one face (square) of the cube described above by its diagonal,  and the height being the length of one edge of the cube. This shape has four and two sharp vertices with solid angle $\pi/4$, $\pi/2$, respectively, and two and seven edges with dihedral angle $\pi/4$, $\pi/2$, respectively. For the system of this configuration, we follow the same process as before, giving the sum of the related corner functions and edge terms, and using values of $\beta(\pi/2)$ and $e(\pi/2)$ and their SDs in the table \ref{edge_corner1}, we obtain corner functions $\beta(\pi/4)$ and edge term coefficients $e(\pi/4)$ and their SDs as shown in the table \ref{edge_corner2}. 
\begin{table}[h]
\caption{The corner functions $\beta(\pi/4)$ and Edge term coefficient $e(\pi/4)$}
\label{edge_corner2}
\hspace{4cm}
\begin{tabular}{|c|ll|}
 \hline
 & $\beta(\pi/4)\pm \beta(\pi/4)_{sd}$ & $e(\pi/4)\pm e(\pi/4)_{sd}$ \\
 \hline 
PBC &  $0.0115 \pm  0.0002$ & $0.1416 \pm 0.0001$\\
FBC &  $0.0114 \pm 0.0001$ & $0.1417 \pm 0.0001$ \\
\hline
\end{tabular}
\end{table} 
The values of corner functions, the edge term coefficients are consistent within the range of their SDs between two cases where the universal lattice satisfies PBCs and FBCs, and both of them are stable values with relative SDs below 0.02 or 0.001. In addition, the values of $b(\pi/4)$ and $e(\pi/4)$ are about twice as large as those of $b(\pi/2)$ and $e(\pi/2)$, respectively. These characteristics are consistent with the fact that the geometric discontinuity of the boundary surface becomes stronger as the solid angle or dihedral angle approaches an acute angle, and the contribution from sharp corners and edges increases, as was the case with the two dimensional lattice.  
\section{conclusions}
We have studied the logarithmic negativity(LN) and mutual information(MI) between two adjacent subsets $A_1$ and $A_2$ in a isolated universal set $U$ of harmonic oscillators on a two dimensional lattice, whose continuum limit is the massless scalar field. We have let the sites of $A_2$ be enclosed by sites of $A_1$ and let the boundary of $A_2$ be one of various shapes with corners whose angles are $\pi/4$, $\pi/2$, $3\pi/4$. Varying the scales of subsystems $A=A_1+A_2$ and $A_2$ kept as similar as possible in their shapes, we have obtained LNs, MIs and from them so-called corner functions for the above corner angles to verify the results of the previous study which treats corner functions mainly from LNs for the case where the canonical variables obey periodic boundary conditions(PBCs).
In the case that canonical variables on the lattice obey PBCs, all the values of corner functions $b(\theta)$ obtained from LNs almost agree with those in the previous studies, although some of the differences are quite large relative to the standard deviations. As for the corner function values $\tilde{b}(\theta)$ obtained from MIs, the values of the corner function $\tilde{b}(\pi/2)$ and $\tilde{b}(\pi/4)$ are almost agreed with those obtained from data of $LN$s. On the other hand, as for the values of $\tilde{b}(3\pi/4)$, considering the magnitude of the SDs and their ratio to those obtained from $LN$s data, or the previous study\cite{Nobili}, they are not stable and not enough consistent to those of the previous study.

We have also examined the values of each corner function in the same way as above, when the canonical variables satisfy the fixed boundary conditions(FBCs), where subsets $A$ and $A_2$ have been taken at various locations in the lattice $U$ to investigate the effect of the fixed ends. The variation of each corner function depending on the location of the region $D$ is shown in Fig.\ref{graph_bs}. In the case that subsets $A$ and $A_2$ are placed near the center of a sufficiently large U, the corner function values and their SDs obtained from LN and MI, respectively, almost agree with those obtained in the $U$ under PBCs. However, assuming the size $l_A$ of the largest $A$, when these subsets approach about 3$l_A$ from the edge of $U$, the values of each corner function starts to deviate from those calculated near the center of $U$. When they approach about $l_A$ from the edge, the deviations increase and the behaviors are unstable. 

Furthermore, We have computed the Renyi entropies of order 1/2 for a three-dimensional lattice, considering, as a subsystem, a set of sites inside a cube varying the scale of the subsystem. By fitting the data with equation (\ref{3DEE}), we have obtained the corner functions and edge term coefficients as shown in table \ref{edge_corner1}. We have compared the SDs of the corner functions with those obtained by fitting with (\ref{3DEE}) minus the edge terms . The result shows that the SDs of the latter are  5 to 10 times larger than that of the former. This fact is the evidence of the existence of the edge terms in the expansions of the Renyi entropy between two adjacent systems whose boundary is a three-dimensional polyhedron. In addition, the Renyi entropies also have been obtained for the sets of sites in triangular prisms with vertices and edges having solid angles and dihedral angles of $\pi/2 $, $\pi/4$ as the subsystem $A$. Using the results of table \ref{edge_corner1}, the corner function $\beta(\pi/4)$, the edge term coefficient $e(\pi/4)$ and their SDs have been obtained as shown in the table \ref{edge_corner2}. 

As in the case of a two-dimensional lattice, the values of the corner functions and edge term coefficients in tables \ref{edge_corner1} and \ref{edge_corner2}, where U satisfies FBCs and the $A$ system is sufficiently distant from the fixed ends of the universal system $U$, are almost identical to the values when $U$ satisfies PBCs. In addition, these values satisfy the stability and monotonically decreasing properties with respect to the solid angle and dihedral angle, respectively, that the corner functions and edge terms coefficients would have.
\newpage\noindent

\end{document}